\documentclass[aps,pre,showkeys,a4paper,floatfix,amsmath,amssymb,superscriptaddress,showpacs]{revtex4}
\usepackage{natbib}
\usepackage{dcolumn}
\usepackage{graphicx}
\usepackage{epsfig}

\begin{document}
\title{Massive Parallel Quantum Computer Simulator}
\author{K. De Raedt}
\affiliation{Department of Computer Science,
University of Groningen, Blauwborgje 3,
NL-9747 AC Groningen, The Netherlands
}
\author{K. Michielsen}
\affiliation{Department of Applied Physics,
Materials Science Centre, University of Groningen, Nijenborgh 4,
NL-9747 AG Groningen, The Netherlands
}
\author{H. De Raedt\footnote{Corresponding author}}
\email{h.a.de.raedt@rug.nl}
\homepage{http://www.compphys.net}
\affiliation{Department of Applied Physics,
Materials Science Centre, University of Groningen, Nijenborgh 4,
NL-9747 AG Groningen, The Netherlands
}
\author{B. Trieu}
\email{b.trieu@fz-juelich.de}
\affiliation{Zentralinstitut f\"ur Angewandte Mathematik,
Forschungszentrum J\"ulich, D-52425 J\"ulich, Germany
}
\author{G. Arnold}
\email{g.arnold@fz-juelich.de}
\affiliation{Zentralinstitut f\"ur Angewandte Mathematik,
Forschungszentrum J\"ulich, D-52425 J\"ulich, Germany
}
\author{M. Richter}
\email{m.richter@fz-juelich.de}
\affiliation{Zentralinstitut f\"ur Angewandte Mathematik,
Forschungszentrum J\"ulich, D-52425 J\"ulich, Germany
}
\author{Th. Lippert}
\email{th.lippert@fz-juelich.de}
\affiliation{Zentralinstitut f\"ur Angewandte Mathematik,
Forschungszentrum J\"ulich, D-52425 J\"ulich, Germany
}
\author{H. Watanabe}
\email{hwatanabe@is.nagoya-u.ac.jp}
\affiliation {Department of Complex Systems Science,
Graduate School of Information Science, Nagoya University,
Furouchou, Chikusaku, Nagoya 464-8601, Japan
}
\author{N. Ito}
\email{ito@ap.t.u-tokyo.ac.jp}
\affiliation{Department of Applied Physics, School of Engineering, The University of Tokyo,
Hongo 7-3-1, Bunkyo-ku, Tokyo 113-8656, Japan
}
\begin{abstract}
We describe portable software to simulate universal quantum computers on massive parallel computers.
We illustrate the use of the simulation software by running various quantum algorithms
on different computer architectures, such as a IBM BlueGene/L, a IBM Regatta p690+, a Hitachi SR11000/J1,
a Cray X1E, a SGI Altix 3700 and clusters of PCs running Windows XP.
We study the performance of the software by simulating quantum computers containing up to 36 qubits,
using up to 4096 processors and up to 1 TB of memory.
Our results demonstrate that the simulator exhibits nearly ideal scaling
as a function of the number of processors and suggest that the simulation software
described in this paper may also serve as benchmark for testing high-end parallel computers.

\end{abstract}
\date{\today}

\maketitle

\def\ORDER#1{\hbox{${\cal O}(#1)$}}
\def\BRA#1{\langle #1 \vert}
\def\KET#1{\vert #1 \rangle}
\def\EXPECT#1{\langle #1 \rangle}
\def\BRACKET#1#2{\langle #1 \vert #2 \rangle}
\def\hbar{{\mathchar'26\mskip-9muh}}
\def\mod{{\mathop{\hbox{ mod }}}}
\def\CNOT{{\mathop{\hbox{CNOT}}}}
\def\Tr{{\mathop{\hbox{Tr}}}}
\def\bPsi{{\mathbf{\Psi}}}
\def\bPhi{{\mathbf{\Phi}}}
\def\bzero{{\mathbf{0}}}
\def\Eq#1{(\ref{#1})}
\def\NOBAR#1{#1}
\def\BAR#1{\overline{#1}}

\section{Introduction}\label{intro}

In this paper, we describe a Fortran 90 software package to simulate universal quantum computers~\cite{NIEL00}.
The software runs on various computer architectures, ranging from PCs to high-end (vector) parallel machines.
The simulator can perform all the quantum operations that are necessary for universal quantum computation.
The maximum number of qubits is set by the memory of the machine on which the code runs.
We present simulation results for quantum computers containing up to 36 qubits.
In view of the fact that the simulation of quantum systems, such as quantum computers,
requires computational resources that grow exponentially with the system size,
this represents a significant advance beyond the state of the art, which is currently around 32 qubits~\cite{FRAU05}.
An overview of quantum computer simulator software is given in Ref.~\cite{RAED05}.

The main focus of the present work is on the design of portable, efficient parallel
simulation code for a universal quantum computer.
A subsequent paper~\cite{TRIE06} will explore optimization techniques to further improve the performance of the parallel code.
In principle, because of this ``universality'', this code for the ideal quantum computer can also be used to simulate
physical systems, such as quantum spin models or models for physical realizations of quantum computers,
by writing the time evolution of the physical
system as a sequence of elementary quantum gate operations~\cite{NIEL00}.
However, this approach is bound to be computationally inefficient for all but nontrivial physical systems.
Instead, it is much more effective to allow for additional unitary transformations
that implement operations such as the time evolution under a Heisenberg Hamiltonian
in an optimal manner~\cite{RAED00,QCEDOWNLOAD,RAED02,RAED05}.
Such an extension, which does not affect the intrinsic performance of the code,
will be dealt with in a future publication.

The paper is organized as follows.
In Section~\ref{QC}, we briefly recall some basic elements of quantum computation.
In Section~\ref{PARA}, we present a scheme
to parallelize the software to simulate
a quantum computer on a massive parallel computer.
We employ the standard Message Passing Interface (MPI)~\cite{MPI} to implement this scheme.
Section~\ref{GUI} describes the main features of the simulator software.
In Section~\ref{RESULTS},
we discuss a few nontrivial quantum algorithms, such as an adder
of two to five registers of qubits and Shor's factoring algorithm,
that are used to validate and benchmark the simulation software.
Section~\ref{BENCHMARKS} presents our benchmark results
of running the quantum algorithms on various parallel computers.
A discussion is given in Section~\ref{SUMM}.

\section{Quantum Computation}\label{QC}

For convenience of the reader, to make the paper self-contained
and to explain the terminology, we summarize some basic aspects of quantum computation.

\subsection{Quantum computer terminology}

The state $\KET{\Phi}$ of a qubit, the elementary storage unit of a quantum computer,
is described by a two-dimensional vector
of Euclidean length one:
\begin{equation}
\KET{\Phi}=a(0)\KET{0}+a(1)\KET{1}=a_0\KET{0} +a_1\KET{1},
\label{STAT0}
\end{equation}
where $\KET{0}$ and $\KET{1}$ denote two orthogonal basis vectors of the
two-dimensional vector space and $a_0\equiv a(0)$ and $a_1\equiv a(1)$ are complex numbers
such that $|a_0|^2+|a_1|^2=1$.
The result of inquiring about the state of a single qubit, that is the outcome of a measurement,
is either 0 or 1.
The frequency of obtaining 0 (1) can be estimated
by repeated measurement of the same state of the
qubit and is given by $|a_0|^2$ ($|a_1|^2$)~\cite{SCHI68,BAYM74,BALL03}.

The internal state of a quantum computer
with $L$ qubits is described by a vector, also called the state vector, in a $D=2^{L}$
dimensional space~\cite{NIEL00}.
Adopting the convention of quantum computation literature~\cite{NIEL00},
the state of an $L$-qubit quantum computer is represented by
\begin{eqnarray}
\KET{\Phi}&=&a({0\ldots00}) \KET{0\ldots00}
+a({0\ldots01}) \KET{0\ldots01}
+\ldots
+a({1\ldots10}) \KET{1\ldots10}
+a({1\ldots11}) \KET{1\ldots11}
\nonumber \\
&=&a_0 \KET{0}
+a_1 \KET{1}
+\ldots
+a_{2^L-2} \KET{2^L-2}
+a_{2^L-1} \KET{2^L-1}
,
\label{STAT2}
\end{eqnarray}
where in the last line of Eq.~\Eq{STAT2}, the binary representation of the integers
$0,\ldots,2^{L-1}$ was used to denote
$\KET{0}\equiv\KET{0\ldots00},\ldots,\KET{2^L-1}\equiv\KET{1\ldots11}$ and
$a_0\equiv a({0\ldots00}),\ldots,a_{2^L-1}\equiv a({1\ldots11})$.
We normalize the state vector, that is $\BRACKET{\Phi}{\Phi}=1$,
by rescaling the complex-valued amplitudes $a_{i}$ according to
\begin{eqnarray}
\sum_{i=0}^{2^L-1}|a_i|^2=1
.
\label{STAT1}
\end{eqnarray}
Note that in this notation it is convenient to number the qubits from 0 to $L-1$, that is
qubit 0 corresponds to the least significant bit of the integer index
that runs from zero to $2^{L-1}$.

A quantum algorithm is a sequence of unitary operations on the vector $\KET{\Phi}$.
It has been shown that an arbitrary unitary operation
can be written as a sequence of single qubit operations
and the CNOT operation on two qubits~\cite{DIVI95a,NIEL00}.
Therefore, single-qubit operations and the CNOT operation
are sufficient to construct a universal quantum computer~\cite{NIEL00}.
We call these operations elementary unitary transformations.
According to quantum theory, after executing a quantum algorithm,
the probability for observing the quantum computer in one of its $2^{L}$
states is given by the square of the absolute value of the
corresponding element of the state vector.

In the quantum computation literature,
the convention is to count each elementary, unitary transformation
as one operation on a quantum computer~\cite{NIEL00}.
However, carrying out a unitary operation on a conventional computer
requires more than one arithmetic operation and it is customary to determine the performance of an algorithm
by counting the number of arithmetic operations.
Although the difference between these two ways of expressing the performance of an algorithm should
be clear from the context, the reader should keep this difference in mind.

\subsection{Single-qubit operations}\label{SQO}

\setlength{\unitlength}{1cm}
\begin{figure*}[t]
\begin{center}
\includegraphics[width=12cm]{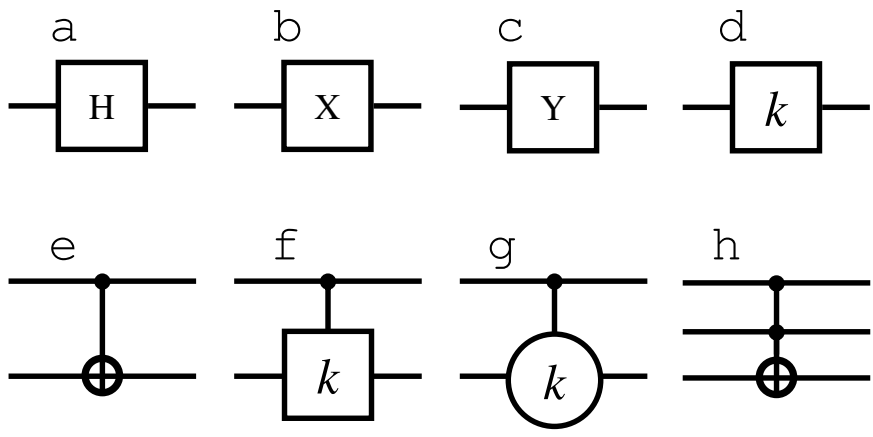}
\end{center}
\caption{
Graphical representation of some of the basic gates used in quantum computation;
(a) Hadamard gate;
(b) Rotation by $\pi/2$ about the $x$-axis;
(c) Rotation by $\pi/2$ about the $y$-axis;
(d) Single qubit phase shift by $\phi =2\pi /2^k$;
(e) CNOT gate;
(f) Controlled phase shift by $\phi =2\pi /2^k$;
(g) Controlled $V$ operation by $\phi =2\pi /2^k$;
(h) Toffoli gate.
The horizontal lines denote the qubits involved in the quantum operations.
The dots and crosses denote the control and target qubits, respectively.}
\label{fig:Gates}
\end{figure*}

Single-qubit operations that are often used in quantum computation
are the Hadamard operation $H$ and rotations $X$ and $Y$ of the state vector
by $\pi/2$ about the $x$ and $y$-axis of the spin-1/2 operator ${\bf S}=(S^x,S^y,S^z)$, representing the qubit.
The Hadamard operation is defined by~\cite{NIEL00}
%
\begin{eqnarray}
H\KET{\Phi}&=&H(a_0\KET{0}+a_1\KET{1})\nonumber\\
&=&\frac{1}{\sqrt{2}}(a_0+a_1)\KET{0}+\frac{1}{\sqrt{2}}(a_0-a_1)\KET{1},
\label{Hadamard}
\end{eqnarray}
the $X$ operation by
\begin{eqnarray}
X\KET{\Phi}&=&e^{i\pi S^x/2}(a_0\KET{0}+a_1\KET{1})\nonumber\\
&=&\frac{1}{\sqrt{2}}(a_0+ia_1)\KET{0}+\frac{1}{\sqrt{2}}(a_1+ia_0)\KET{1},
\label{Xoperation}
\end{eqnarray}
and the $Y$ operation by
\begin{eqnarray}
Y\KET{\Phi}&=&e^{i\pi S^y/2}(a_0\KET{0}+a_1\KET{1})\nonumber\\
&=&\frac{1}{\sqrt{2}}(a_0+a_1)\KET{0}+\frac{1}{\sqrt{2}}(a_1-a_0)\KET{1}.
\label{Yoperation}
\end{eqnarray}
The phase shift operation $R(\phi)$ is defined by
\begin{eqnarray}
R(\phi )\KET{\Phi}&=&e^{i\phi/2}e^{-i\phi S^z}(a_0\KET{0}+a_1\KET{1})\nonumber\\
&=&a_0\KET{0}+a_1e^{i\phi}\KET{1},
\label{singlephase}
\end{eqnarray}
%
$R(\phi)$ changes the phase of the amplitude of the $\KET{1}$ component of the state vector only.
Quantum algorithms are often represented by quantum networks, diagrams that show the order of the
operations performed on the qubits.
The graphical symbols of $H$, $X$, $Y$, and $R(\phi =2\pi/2^k)$ are shown in Fig.~\ref{fig:Gates}.
The inverse and transpose of a unitary operation $U$ are denoted by $\BAR U$ and $U^T$, respectively.

\subsection{Two-qubit operations: CNOT and controlled phase shift}\label{CNOT}

Computation requires some form of communication between the qubits.
Any form of communication
between qubits can be reduced to a combination of single-qubit operations
and the CNOT operation, a two-qubit operation~\cite{DIVI95a,NIEL00}.
By definition, the CNOT gate flips the target qubit if
the control qubit is in the state $\KET{1}$~\cite{NIEL00}.
If we take qubit zero
(that is the least significant bit in the binary notation of an integer)
as the control qubit and qubit one as the target qubit then we have
\begin{eqnarray}
\CNOT_{10}\KET{\Phi}&=&\CNOT_{10}
(a_0\KET{00}+a_1\KET{01}+a_2\KET{10}+a_3\KET{11})
\nonumber\\
&=&a_0\KET{00}+a_3\KET{01}+a_2\KET{10}+a_1\KET{11}
\nonumber\\
&=&a_0\KET{0}+a_3\KET{1}+a_2\KET{2}+a_1\KET{3}.
\label{CNOT2}
\end{eqnarray}
The graphical symbol of the CNOT operation is
shown in Fig.~\ref{fig:Gates}e.
The dot (cross) denotes the control (target) qubit.

Another frequently used operation is the controlled phase shift.
The controlled phase shift operation with control qubit 0
and target qubit 1 is defined by
\begin{eqnarray}
R_{10}(\phi)\KET{\Phi}&=&
R_{10}(\phi)(a_0\KET{00}+a_1\KET{01}+a_2\KET{10}+a_3\KET{11}),
\nonumber\\
&=&a_0\KET{00}+a_1\KET{01}+a_2\KET{10}+e^{i\phi}a_3\KET{11}
\label{COMM5}
.
\end{eqnarray}
Graphically, the controlled phase shift $R_{ji}(\phi=2\pi/2^k)$
is represented by a vertical line connecting a dot (control qubit)
and a box denoting a single qubit phase shift by $2\pi /2^k$ (see Fig.~\ref{fig:Gates}f).

The CNOT operation is a special case of the controlled
unitary transformation $V$.
If qubit zero is the control qubit and qubit one is the target qubit then
\begin{eqnarray}
V_{10}(\phi)\KET{\Phi}&=&
V_{10}(\phi)(a_0\KET{00}+a_1\KET{01}+a_2\KET{10}+a_3\KET{11}),
\nonumber\\
&=&a_0\KET{00}+\frac{(1+e^{i\phi})a_1+(1-e^{i\phi})a_3}{2}\KET{01}
+a_2\KET{10}+
\frac{(1-e^{i\phi})a_1+(1+e^{i\phi})a_3}{2}\KET{11}
\label{COMM4}
.
\end{eqnarray}
The graphical symbol of the controlled $V$ operation, $V_{ji}(\phi=2\pi/2^k)$, is
a vertical line connecting a dot
and a circle containing the value $k$ (see Fig.~\ref{fig:Gates}g).

\subsection{Three-qubit operation: Toffoli gate}\label{TOFF}

The Toffoli gate is a generalization of the CNOT gate
in the sense that it has two control qubits and one target qubit~\cite{BARE95,NIEL00}.
The target qubit flips if and only if the two control qubits are set.
If we take qubit zero and qubit one
as the control qubits and qubit two as the target qubit then we have
\begin{eqnarray}
T_{210}\KET{\Phi}&=&T_{210}
(a_0\KET{000}+a_1\KET{001}+a_2\KET{010}+a_3\KET{011}+a_4\KET{100}+a_5\KET{101}+a_6\KET{110}+a_7\KET{111})
\nonumber\\
&=&a_0\KET{000}+a_1\KET{001}+a_2\KET{010}+a_7\KET{011}+a_4\KET{100}+a_5\KET{101}+a_6\KET{110}+a_3\KET{111}
\nonumber\\
&=&a_0\KET{0}+a_1\KET{1}+a_2\KET{2}+a_7\KET{3}+a_4\KET{4}+a_5\KET{5}+a_6\KET{6}+a_3\KET{7}.
\label{Toffoli}
\end{eqnarray}
Symbolically the Toffoli gate is represented by a vertical line
connecting two dots (control qubits) and
one cross (target qubit), as shown in Fig.~\ref{fig:Gates}h.

\section{Parallelization}\label{PARA}

\subsection{General computational aspects}

Computer memory and CPU time put limitations on the size of the quantum computer that can be simulated on
a conventional digital computer.
The required CPU time is mainly determined by the number of operations to be performed on the qubits.
The CPU time does not put a hard limit on the simulation. However, the memory of the computer does.
According to Eq.~\Eq{STAT2},
the state of a $L$-qubit quantum computer is represented by a complex-valued vector of length $D = 2^L$.
In view of the potentially large number of arithmetic operations,
it is advisable to use 13 - 15 digit floating-point arithmetic (corresponding
to 8 bytes for a real number).
Thus, to represent a state of the quantum system of $L$ qubits
in a conventional, digital computer, we need at least $2^{L+4}$ bytes.
Hence, the amount of memory that is required to simulate a quantum
computer with $L$ qubits increases exponentially with the number of qubits $L$.
For example, for $L=24$ ($L=36$)
we need at least 256 MB (1TB) of memory to store a single arbitrary state $\KET{\Phi}$.

As seen in Section~\ref{QC}, operations $U$ on the state vector $\KET{\Phi}$ result in a transformation of
the amplitudes of the basis states in $\KET{\Phi}$.
More specifically, let us denote
\begin{eqnarray}
\KET{\Phi}=
& &a({00\ldots 0}) \KET{00\ldots 0}
+a({0\ldots 01}) \KET{0\ldots 01}
+\ldots 
+a({01\ldots 1}) \KET{01\ldots 1}
+a({11\ldots 1}) \KET{11\ldots 1}
,
\label{NUM0}
\end{eqnarray}
and
\begin{eqnarray}
\KET{\Phi'}&=&U\KET{\Phi}\nonumber \\
&=&a'({00\ldots 0}) \KET{00\ldots 0}
+a'({0\ldots 01}) \KET{0\ldots 01}
+\ldots 
+a'({01\ldots 1}) \KET{01\ldots 1}
+a'({11\ldots 1}) \KET{11\ldots 1}
.
\label{NUM1}
\end{eqnarray}

We first consider the single-qubit operations on qubit $j$ that transform $\KET{\Phi}$ in $\KET{\Phi'}=U_j\KET{\Phi}$.
From Eq.~\Eq{Hadamard}, it follows that the Hadamard operation on qubit $j$, $H_j$,
transforms the amplitudes according to
\begin{eqnarray}
a'({*\ldots*0_j*\ldots*})&=&\frac{1}{\sqrt{2}}(a({*\ldots*0_j*\ldots*})+a({*\ldots*1_j*\ldots*}))\nonumber \\
a'({*\ldots*1_j*\ldots*})&=&\frac{1}{\sqrt{2}}(a({*\ldots*0_j*\ldots*})-a({*\ldots*1_j*\ldots*})),
\label{NUM2}
\end{eqnarray}
where we use the asterisk to indicate that the bits on the corresponding positions
are the same.
From Eq.~\Eq{Xoperation}, it follows that for $X_j$ operating on $\KET{\Phi}$
the elements of $\KET{\Phi'}$ are obtained
by
\begin{eqnarray}
a'({*\ldots*0_j*\ldots*})&=&\frac{1}{\sqrt{2}}(a({*\ldots*0_j*\ldots*})+ia({*\ldots*1_j*\ldots*}))\nonumber \\
a'({*\ldots*1_j*\ldots*})&=&\frac{1}{\sqrt{2}}(a({*\ldots*1_j*\ldots*})+ia({*\ldots*0_j*\ldots*})).
\label{NUM3}
\end{eqnarray}
In the case of $\KET{\Phi'}=Y_j\KET{\Phi}$, it follows from Eq.~\Eq{Yoperation} that
\begin{eqnarray}
a'({*\ldots*0_j*\ldots*})&=&\frac{1}{\sqrt{2}}(a({*\ldots*0_j*\ldots*})+a({*\ldots*1_j*\ldots*}))\nonumber \\
a'({*\ldots*1_j*\ldots*})&=&\frac{1}{\sqrt{2}}(a({*\ldots*1_j*\ldots*})-a({*\ldots*0_j*\ldots*})).
\label{NUM4}
\end{eqnarray}
From Eq.~\Eq{singlephase} it follows that we obtain $\KET{\Phi'}=R_j(\phi )\KET{\Phi}$ by leaving the amplitudes,
for which the $j$th bit of their index is zero, unchanged and by multiplying the amplitudes, for which the $j$th bit
of their index is one, with the phase factor $e^{i\phi}$. Hence we have,
\begin{eqnarray}
a'({*\ldots*0_j*\ldots*})&=&a({*\ldots*0_j*\ldots*})\nonumber \\
a'({*\ldots*1_j*\ldots*})&=&e^{i\phi}a({*\ldots*1_j*\ldots*}).
\label{NUM5}
\end{eqnarray}
In summary, performing an operation on qubit $j$ requires in general an update of $2^{L}$
elements of $\KET{\Phi}$. The single-qubit phase shift operation forms an exception and requires an update
of $2^{L-1}$ single amplitudes only.
Note that all these operations can be done in place,
that is, without using another vector of length $2^L$.

We now consider two-qubit operations $\KET{\Phi'}=U_{kj}\KET{\Phi}$, with $j<k$.
For the CNOT operation $\CNOT_{kj}$, where qubit $j$ is the control qubit and qubit $k$ is the target qubit,
amplitudes for which bit $j$ of their index is one and bit $k$ of their index is zero need to be swapped with
amplitudes for which bits $j$ and $k$ of their index are one (see Eq.~\Eq{CNOT2}).
We have
\begin{eqnarray}
a'({*\ldots*0_k*\ldots*0_j*\ldots*})&=&a({*\ldots*0_k*\ldots*0_j*\ldots*})\nonumber \\
a'({*\ldots*0_k*\ldots*1_j*\ldots*})&=&a({*\ldots*1_k*\ldots*1_j*\ldots*})\nonumber \\
a'({*\ldots*1_k*\ldots*0_j*\ldots*})&=&a({*\ldots*1_k*\ldots*0_j*\ldots*})\nonumber \\
a'({*\ldots*1_k*\ldots*1_j*\ldots*})&=&a({*\ldots*0_k*\ldots*1_j*\ldots*}).
\label{NUM6}
\end{eqnarray}
For the controlled-$V$ operation $V_{kj}$, it follows from Eq.~\Eq{COMM4} that the amplitudes change according to the
rules
\begin{eqnarray}
a'({*\ldots*0_k*\ldots*0_j*\ldots*})&=&a({*\ldots*0_k*\ldots*0_j*\ldots*})\nonumber \\
a'({*\ldots*0_k*\ldots*1_j*\ldots*})&=&\frac{1}{2}\left[(1+e^{i\phi})a({*\ldots*0_k*\ldots*1_j*\ldots*})
+(1-e^{i\phi})a({*\ldots*1_k*\ldots*1_j*\ldots*})\right]\nonumber \\
a'({*\ldots*1_k*\ldots*0_j*\ldots*})&=&a({*\ldots*1_k*\ldots*0_j*\ldots*})\nonumber \\
a'({*\ldots*1_k*\ldots*1_j*\ldots*})&=&\frac{1}{2}\left[(1-e^{i\phi})a({*\ldots*0_k*\ldots*1_j*\ldots*})
+(1+e^{i\phi})a({*\ldots*1_k*\ldots*1_j*\ldots*})\right].
\label{NUM7}
\end{eqnarray}
Thus, performing two-qubit operations such as the CNOT and the controlled-$V$
amount to update $2^{L-1}$ amplitudes.

For the Toffoli gate $T_{lkj}$ ($j<k<l$), where qubits $j$ and $k$ are the control qubits and qubit $l$ is
the target qubit, only $2^{L-2}$ elements of $\KET{\Phi}$ need to be updated,
as can be seen from Eq.~\Eq{Toffoli}. The update rules read
\begin{eqnarray}
a'({*\ldots*0_l*\ldots*1_k*\ldots*1_j*\ldots*})&=&a({*\ldots*1_l*\ldots*1_k*\ldots*1_j*\ldots*})\nonumber \\
a'({*\ldots*1_l*\ldots*1_k*\ldots*1_j*\ldots*})&=&a({*\ldots*0_l*\ldots*1_k*\ldots*1_j*\ldots*}),
\label{NUM8}
\end{eqnarray}
and the other amplitudes remain unchanged.

All the qubit operations discussed here
can be carried out in \ORDER{2^L} floating-point operations
(here and in the sequel, if we count operations on a conventional
computer, we make no distiction between operations such as add, multiply,
or get from and put to memory).
The qubit operations described earlier
suffice to implement any unitary transformation on the state vector.
As a matter of fact, the simple rules described are all that we need to
simulate a universal quantum computer on a conventional computer.

The fact that both the amount of memory and
arithmetic operations increase exponentially with the number of qubits
is the major bottleneck for simulating quantum computers (or quantum systems in general)
on a conventional computer.
However, in contrast to the CPU time, the total amount of memory that is available on a computer
puts a hard limit on the simulations that can be performed.
From the user's perspective, the memory on a computer can be shared or
distributed. On a shared memory computer, the state $\KET{\Phi}$
of the quantum computer can be completely stored in the memory and all processors can access the entire memory.
On a distributed memory machine, the elements of $\KET{\Phi}$ are physically distributed over
different nodes and each processor has direct access to its own local memory only.
In the latter case, some extra programming is required to perform the communication between
the processors.
We use the standard Message Passing Interface (MPI) to perform the data communication~\cite{MPI}.
We assume that the reader has some basic knowledge of MPI programming~\cite{MPI}.

\subsection{Implementation}\label{PARA1}

Each MPI process has a rank and a local memory. We assume that the local memory
can store the $2^M$ amplitudes of the basis states of $M$ qubits. Hence, to simulate a $L$-qubit quantum computer
we need $N=2^L/2^M$ MPI processes.
From Eq.~\Eq{STAT2}, it is clear that the amplitudes $a(x_{L-1} \ldots x_0)$
where $x_i=0,1$ for $i=0,\ldots,L-1$,
can be stored at the local memory address $A=\sum_{i=0}^{M-1} 2^ix_i$ of the MPI process
with rank $R=\sum_{i=M}^{L-1} 2^{i-M} x_i$. In binary notation the local memory address
and the rank of the processor reads $A=(x_{M-1}\ldots x_0)$ and $R=(x_{L-1}\ldots x_M)$, respectively.
Recall that the qubits are numbered from 0 to $L-1$, that is
qubit 0 corresponds to the least significant bit of the integer index, running from zero to $2^{L-1}$, of the amplitude.
An example for $L=4$ and $M=2$ is shown in the first four columns of Table~\ref{onequbittab}.
The MPI processes are labeled with their ranks running from ${\bf 00}$ to ${\bf 11}$.
The memory addresses are also running from ${\it 00}$ to ${\it 11}$ for this example.

\begin{table}[t]
\caption{ Single-qubit operation:
Distribution of the amplitudes $a(x_{L-1} \ldots x_0)$ over the $N$ MPI processes with $2^M$ memory locations
after application of the permutation $\sigma_p$, for the case $L=4$ and $M=2$ ($N=2^L/2^M$).
$\sigma_1$ correponds to the identity permutation, $\sigma_2$ interchanges local qubit 0 and nonlocal qubit 2,
$\sigma_3$ interchanges local qubit 2 and nonlocal qubit 3, after having interchanged qubits 0 and 2.
The rank $R$ of the MPI process is shown (in binary notation) in the second row.
The local memory addresses are shown in the first column.
}
\begin{center}
\begin{tabular}{|c||c|c|c|c||c|c|c|c||c|c|c|c||}
\hline
&\multicolumn{4}{c||}{$\sigma_1=\left(\begin{array}{cccc}3&2&1&0 \\3&2&1&0\end{array}\right)$}
&\multicolumn{4}{c||}{$\sigma_2=\left(\begin{array}{cccc}3&2&1&0 \\3&0&1&2\end{array}\right)$}
&\multicolumn{4}{c||}{$\sigma_3=\left(\begin{array}{cccc}3&2&1&0 \\2&0&1&3\end{array}\right)$}
\\
\hline
& {\bf 00} & {\bf 01} & {\bf 10} & {\bf 11}  & {\bf 00} & {\bf 01} & {\bf 10} & {\bf 11}& {\bf 00} & {\bf 01} & {\bf 10} & {\bf 11} \\
\hline
 {\it 00}  &  a(0000)  &  a(0100)  &  a(1000)  & a(1100)    &  a(0000)  &  a(0001)  &  a(1000)  & a(1001)  &  a(0000)  &  a(0001)  &  a(0100)  & a(0101)\\
\hline
 {\it 01}  &  a(0001)  &  a(0101)  &  a(1001)  & a(1101)    &  a(0100)  &  a(0101)  &  a(1100)  & a(1101)  &  a(1000)  &  a(1001)  &  a(1100)  & a(1101)\\
\hline
 {\it 10}  &  a(0010)  &  a(0110)  &  a(1010)  & a(1110)    &  a(0010)  &  a(0011)  &  a(1010)  & a(1011)  &  a(0010)  &  a(0011)  &  a(0110)  & a(0111)\\
\hline
 {\it 11}  &  a(0011)  &  a(0111)  &  a(1011)  & a(1111)    &  a(0110)  &  a(0111)  &  a(1110)  & a(1111)  &  a(1010)  &  a(1011)  &  a(1110)  & a(1111)\\
\hline
\end{tabular}
\label{onequbittab}
\end{center}
\end{table}

As can be seen from Eqs.~\Eq{NUM2}-\Eq{NUM5}, performing an operation on qubit $j$ requires in general
an update of $2^{L}$ elements of $\KET{\Phi}$.
A qubit $j$ for which $j<M$  we call a ``local'' qubit.
In the example shown in the first four columns of Table~\ref{onequbittab}
qubits 0 and 1 are local qubits.
Updating the amplitudes is easy if the qubit is local because this requires no communication between the
MPI processes.

Similarly, if $j\geq M$ we call qubit $j$ a ``nonlocal'' qubit.
An operation on qubit $j$ requires communication between the MPI processes.
In the example shown in the first four columns of Table~\ref{onequbittab}, qubits 2 and 3 are nonlocal.
Note that the nonlocal qubits form the binary representation of the rank of the
corresponding MPI process and that the local qubits form the binary representation
of the memory address.
An obvious way of communication would be for pairs of MPI processes
to first interchange one half of their data.
Then, the operations on the amplitudes can be performed as if the qubit was local.
Finally, the MPI processes need again to interchange the (modified) data.
A clear drawback of this method is that half of the amplitudes needs to be interchanged twice for every
single-qubit operation on qubit $j$ for which $j\ge M$.

In order to reduce the amount of communication between the MPI
processes we use a different method to handle the operation on nonlocal qubits.
We introduce a permutation $\sigma :\{ L-1,\ldots ,0\}\rightarrow \{\sigma (L-1),\ldots ,\sigma (0)\}$
of the $L$ qubits.
We denote the permutation by a matrix with two rows and $L$ columns. The first row contains integer numbers ordered
from $L-1$ to 0. These numbers correspond to the position of the bits in the index of the amplitude. The second row also contains
integer numbers ranging from $L-1$ to 0 but they are not necessarily ordered. These numbers refer to the qubits.
Local qubit $k$ corresponds to bit $l=\sigma^{-1}(k)< M$ of the index of the amplitude
and nonlocal qubit $m$ corresponds to bit $n=\sigma^{-1}(m)\ge M$ of the index of the amplitude.
After applying the permutation $\sigma$ the amplitudes $a(x_{L-1} \ldots x_0)$ ($x_i=0,1$) are stored at the address
$A=\sum_{i=0}^{M-1} 2^ix_{\sigma (i)}$ of the local memory assigned
to the MPI process with rank $R=\sum_{i=M}^{L-1} 2^{i-M} x_{\sigma (i)}$.
In binary notation we have
$A=(x_{\sigma (M-1)}\ldots x_{\sigma (0)})$ and $R=(x_{\sigma (L-1)}\ldots x_{\sigma (M)})$.
Note that after applying the permutation $\sigma$,
the nonlocal qubits still form the binary representation of the rank of the
corresponding MPI process.
Similarly, the local qubits still form the binary representation of the address.

Logically, to make a nonlocal qubit $m$ local, all we have to do is to select a permutation
that interchanges a local qubit, say $k$, and the nonlocal qubit $m$ and leaves the other qubits in place.
Clearly, it is always easy to find such a permutation.
We now consider what it actually means for the parallel machine to perform an
interchange of a local and nonlocal qubit. A simple reshuffling of the corresponding amplitudes
would work but is not very efficient: We would like to minimize the amount of interprocess communication.

In terms of data transfer between MPI processes,
a permutation that interchanges local qubit $k$ and nonlocal qubit $m$,
swaps the amplitudes with addresses
\begin{eqnarray}
A=(*\ldots *0_l*\ldots *)\quad &\hbox{\rm of MPI process with rank}&\quad
R=(*\ldots *1_n*\ldots *)\nonumber\\
&\hbox{and}&\nonumber\\
A=(*\ldots *1_l*\ldots *)\quad &\hbox{\rm of MPI process with rank}&\quad
R=(*\ldots *0_n*\ldots *),
\label{onequbitchange}
\end{eqnarray}
where we use the asterisk to indicate that the bits on the corresponding position are the same.
Hence, in this operation, only half of the amplitudes in the
memory of an MPI process are swapped against half of the amplitudes in the memory of another MPI process.
After this swap operation, the previously nonlocal qubit $m$ has become local,
and we can carry out the operations on this qubit in exactly the same way
as we did for the originally local qubits, that is fully parallel.
Using this scheme we don't need to send the modified amplitudes back to the MPI processes from which
they originate as the permutation keeps track of the memory adresses of the amplitudes.

An example of this process of swapping data is shown in Table~\ref{onequbittab}, for $L=4$, $M=2$.
The permutations $\sigma$ that have been applied are shown on the top row.
The four columns below $\sigma_1$ show the original content of the local memories.
In this example, qubits 0 and 1 are local and qubits 2 and 3 are nonlocal.
Hence, operations on qubit 0 or 1 can be performed in parallel by each MPI process.
However, operations on qubit 2 and 3 require communication between the MPI processes.
Let us now assume that we want to carry out an operation on qubit 2.
According to our scheme, this requires that the pairs of amplitudes $a(x_3,0_2,x_1,x_0)$ and $a(x_3,1_2,x_1,x_0)$
reside in the same local memory.
Conceptually, this can be accomplished by rearranging the amplitudes over the
local memories according to the permutation $\sigma_2$ (see Table~\ref{onequbittab})
whereas the actual data exchange is carried out according to Eq.~\Eq{onequbitchange}.

Now qubits 2 and 1 are local and qubits 0 and 3 are nonlocal. Assume that we now want to
operate on qubit 3 which is currently nonlocal.
Then, we may interchange, for example, local qubit 2 with the nonlocal qubit 3.
This can be accomplished by applying the permutation $\sigma_3$ (see Table~\ref{onequbittab})
and data exchange rule Eq.~\Eq{onequbitchange}.
At this point, qubits 3 and 1 are local and qubits 0 and 2 are nonlocal.

To summarize: A single-qubit operation on a nonlocal qubit consists
of a local-nonlocal data exchange defined by Eq.~\Eq{onequbitchange}, followed
by the actual unitary operation.
By construction, there is no interprocess communication during and after the latter.
If the qubit is local, we simply skip the step of exchanging data.
From Eq.~\Eq{NUM5}, it is clear that the single-qubit phase-shift operation
never requires communication between MPI processes.
Thus, for this particular operation there is additional room for optimization.
In our present simulation code, we have choosen not to optimize gate operations
on this level.
Optimization of parallel code for specific gates will be dealt with in depth in a subsequent publication~\cite{TRIE06}.

\begin{table}[t]
\caption{Two-qubit operation:
Distribution of the amplitudes $a(x_{L-1} \ldots x_0)$ over the $N$ MPI processes with $2^M$ memory locations
after application of the permutation $\sigma_p$, for the case $L=4$ and $M=2$ ($N=2^L/2^M$).
$\sigma_1$ correponds to the identity permutation, $\sigma_2$ interchanges local qubit 0 with nonlocal qubit 2,
and local qubit 1 with nonlocal qubit 3.
The rank $R$ of the MPI process is shown (in binary notation) in the second row.
The local memory addresses are shown in the first column.
}
\begin{center}
\begin{tabular}{|c||c|c|c|c||c|c|c|c||}
\hline
&\multicolumn{4}{c||}{$\sigma_1=\left(\begin{array}{cccc}3&2&1&0 \\3&2&1&0\end{array}\right)$}
&\multicolumn{4}{c||}{$\sigma_2=\left(\begin{array}{cccc}3&2&1&0 \\1&0&3&2\end{array}\right)$}
\\
\hline
& {\bf 00} & {\bf 01} & {\bf 10} & {\bf 11}  & {\bf 00} & {\bf 01} & {\bf 10} & {\bf 11}\\
\hline
 {\it 00}  &  a(0000)  &  a(0100)  &  a(1000)  & a(1100)    &  a(0000)  &  a(0001)  &  a(0010)  & a(0011)\\
\hline
 {\it 01}  &  a(0001)  &  a(0101)  &  a(1001)  & a(1101)    &  a(0100)  &  a(0101)  &  a(0110)  & a(0111)\\
\hline
 {\it 10}  &  a(0010)  &  a(0110)  &  a(1010)  & a(1110)    &  a(1000)  &  a(1001)  &  a(1010)  & a(1011)\\
\hline
 {\it 11}  &  a(0011)  &  a(0111)  &  a(1011)  & a(1111)    &  a(1100)  &  a(1101)  &  a(1110)  & a(1111)\\
\hline
\end{tabular}
\label{twoqubittab}
\end{center}
\end{table}

Extending the single-qubit scheme to two and three-qubit operations is conceptually straigthforward but
the rules to determine which processes have to exchange data become more complicated.
From Eqs.~\Eq{NUM6} and \Eq{NUM7}, it follows that a two-qubit operation
involves an update of $2^{L-1}$ elements of $\KET{\Phi}$.
If these elements are located in the same local memory, that is if both qubits are local,
no communication between the MPI processes is required and the operation can be performed
independently by all the MPI processes.
However, if one of the qubits is nonlocal, some communication is necessary. If only one qubit is nonlocal,
we can use the same procedure as described above to interchange the qubit for a local one. If both qubits are nonlocal,
we can also use the same procedure, but twice. For each qubit interchange, half of the amplitudes in the
memory of a MPI process is swapped against half of the amplitudes in the memory of another MPI process.
Hence, this scheme results in a full memory swap.
However, this amount of communication can be reduced by exchanging
two nonlocal qubits and two local qubits simultaneously.
We illustrate this by the example for the case $L=4$, $M=2$, shown in Table~\ref{twoqubittab}.
The four columns below $\sigma_1$ show the original content of the local memory.
In this example, qubits 0 and 1 are local and qubits 2 and 3 are nonlocal.
Hence, two-qubit operations on qubit 0 and  1 can be performed independently by each MPI process.
However, all other two-qubit operations require communication between the MPI processes.

Let us now assume that we want to do a two-qubit operation on qubits 2 and 3.
Permutation $\sigma_2$ interchanges nonlocal qubits 2 and 3 with local qubits 0 and 1, respectively.
From Table~\ref{twoqubittab} we see that each MPI process has to communicate with three other processes.
For example, MPI process ${\bf 00}$ keeps the amplitude in address ${\it 00}$, interchanges its amplitude in
address ${\it 01}$ with the amplitude in address ${\it 00}$ of MPI process ${\bf 01}$,
interchanges its amplitude in
address ${\it 10}$ with the amplitude in address ${\it 00}$ of MPI process ${\bf 10}$,
and interchanges its amplitude in address ${\it 11}$ with the amplitude in address ${\it 00}$
of MPI process ${\bf 11}$.

In general, for a two-qubit interchange, the MPI processes communicate in groups of four (in the example of
Table~\ref{twoqubittab} there is only one group of four).
Each MPI process keeps one quarter of the amplitudes in its local memory and interchanges
the other three quarters of amplitudes with amplitudes in the local memories of the other MPI processes in the group.
Swapping of the amplitudes can be accomplished by applying the permutation
$\sigma :\{ L-1,\ldots ,0\}\rightarrow \{\sigma (L-1),\ldots ,\sigma (0)\}$ of the $L$ qubits so that
the MPI processes interchange local qubits $k_1$, $k_2$ for nonlocal qubits $m_1$, $m_2$.
Local qubit $k_i$ corresponds to bit $l_i=\sigma^{-1}(k_i) < M$ of the index of the amplitude
and nonlocal qubit $m_i$ corresponds to bit $n_i=\sigma^{-1}(m_i)\geq M$ of the index of the amplitude, where $i=1,2$.
Adopting the convention that $l_2>l_1$ and $m_2>m_1$,
interchanging local qubits $k_1$, $k_2$ and nonlocal qubits $m_1$, $m_2$
amounts to swapping the amplitudes with local memory addresses
\begin{eqnarray}
A=(*\ldots *0_{l_2}*\ldots *0_{l_1}*\ldots *)\quad &\hbox{\rm of MPI process with rank}&
\quad R=(*\ldots *0_{n_2}*\ldots *1_{n_1}*\ldots *)\nonumber\\
&\hbox{and}&\nonumber\\
A=(*\ldots *0_{l_2}*\ldots *1_{l_1}*\ldots *)\quad &\hbox{\rm of MPI process with rank}&
\quad R=(*\ldots *0_{n_2}*\ldots *0_{n_1}*\ldots *),
\label{twoqubitchange0}
\end{eqnarray}
\begin{eqnarray}
A=(*\ldots *0_{l_2}*\ldots *0_{l_1}*\ldots *)\quad &\hbox{\rm of MPI process with rank}&
\quad R=(*\ldots *1_{n_2}*\ldots *0_{n_1}*\ldots *)\nonumber\\
&\hbox{and}&\nonumber\\
A=(*\ldots *1_{l_2}*\ldots *0_{l_1}*\ldots *)\quad &\hbox{\rm of MPI process with rank}&
\quad R=(*\ldots *0_{n_2}*\ldots *0_{n_1}*\ldots *),
\label{twoqubitchange1}
\end{eqnarray}
\begin{eqnarray}
A=(*\ldots *0_{l_2}*\ldots *0_{l_1}*\ldots *)\quad &\hbox{\rm of MPI process with rank}&
\quad R=(*\ldots *1_{n_2}*\ldots *1_{n_1}*\ldots *)\nonumber\\
&\hbox{and}&\nonumber\\
A=(*\ldots *1_{l_2}*\ldots *1_{l_1}*\ldots *)\quad &\hbox{\rm of MPI process with rank}&
\quad R=(*\ldots *0_{n_2}*\ldots *0_{n_1}*\ldots *),
\label{twoqubitchange2}
\end{eqnarray}
\begin{eqnarray}
A=(*\ldots *0_{l_2}*\ldots *1_{l_1}*\ldots *)\quad &\hbox{\rm of MPI process with rank}&
\quad R=(*\ldots *1_{n_2}*\ldots *0_{n_1}*\ldots *)\nonumber\\
&\hbox{and}&\nonumber\\
A=(*\ldots *1_{l_2}*\ldots *0_{l_1}*\ldots *)\quad &\hbox{\rm of MPI process with rank}&
\quad R=(*\ldots *0_{n_2}*\ldots *1_{n_1}*\ldots *),
\label{twoqubitchange3}
\end{eqnarray}
\begin{eqnarray}
A=(*\ldots *0_{l_2}*\ldots *1_{l_1}*\ldots *)\quad &\hbox{\rm of MPI process with rank}&
\quad R=(*\ldots *1_{n_2}*\ldots *1_{n_1}*\ldots *)\nonumber\\
&\hbox{and}&\nonumber\\
A=(*\ldots *1_{l_2}*\ldots *1_{l_1}*\ldots *)\quad &\hbox{\rm of MPI process with rank}&
\quad R=(*\ldots *0_{n_2}*\ldots *1_{n_1}*\ldots *),
\label{twoqubitchange4}
\end{eqnarray}
\begin{eqnarray}
A=(*\ldots *1_{l_2}*\ldots *0_{l_1}*\ldots *)\quad &\hbox{\rm of MPI process with rank}&
\quad R=(*\ldots *1_{n_2}*\ldots *1_{n_1}*\ldots *)\nonumber\\
&\hbox{and}\nonumber\\
A=(*\ldots *1_{l_2}*\ldots *1_{l_1}*\ldots *)\quad &\hbox{\rm of MPI process with rank}&
\quad R=(*\ldots *1_{n_2}*\ldots *0_{n_1}*\ldots *).
\label{twoqubitchange5}
\end{eqnarray}
As before, the asterisks in Eqs.~\Eq{twoqubitchange0}-\Eq{twoqubitchange5}
indicate that the bits on the corresponding position are the same.
As in the case of the single-qubit operations,
particular two-qubit operations (such as the controlled phase shift)
allow for additional optimization, but we have choosen not to do so
because we wanted to implement all two-qubit operations in the same manner.

The algorithm to interchange a pair of local qubits with a pair of nonlocal qubits
(see Eqs.~\Eq{twoqubitchange0}-\Eq{twoqubitchange5}) can be generalized to swap
as many local qubits with nonlocal qubits, with the restriction that
the number of qubits to be swapped cannot exceed the number of local or nonlocal qubits.
Exchanging $K$ pairs of local and nonlocal qubits simultaneously requires each MPI process
to send $(2^K-1)2^M/2^K$ amplitudes to another MPI process.
Sequentially exchanging the $K$ pairs amounts to sending $K2^M/2$ amplitudes to another process.
Thus, exchanging $K$ pairs of local and nonlocal qubits simultaneously involves less communication.
However, the more qubits we swap simultaneously,
the larger becomes the group of MPI processes that communicate with each other. If this number becomes too big
the computer might hang in network collisions. So, in practice there is a limit to the choice of $K$.
In our implemention of the algorithm, $K$ is an input variable.
For each MPI process, we first compute the ranks of the other MPI processes with which the MPI process has to communicate.
We do this on the basis of the $K$ pairs of local and nonlocal qubits that need to be swapped.
Before exchanging data between the MPI processes belonging to one group,
we fill a buffer for each MPI process of the group.
The buffer contains all amplitudes that need to be send from
one MPI process to the other MPI processes of the group.
In order to reduce the memory allocation for the buffers, an essential requirement for
simulating quantum computers with a large number of qubits $L$,
we split the MPI send instruction in a fixed (user controlled), smaller number of send instructions.

\section{Simulator}\label{GUI}

We have implemented the algorithm described in Section~\ref{PARA} in Fortran 90. The computer code
contains all quantum operations that are necessary for universal quantum computation and runs on machines with
distributed and/or shared memory.
Standard MPI is used for interprocess communication.
Optionally, OpenMP can be used for parallel processing within each MPI process.
This computer code forms the engine of the parallel quantum computer simulator.
It compiles (without modifications) and runs on various computer architectures,
ranging from PCs to high-end (vector) parallel machines.
The number of qubits in the computer codes is limited to 62, but
in practice the maximum number of qubits is set by the memory of the machine on which the code runs.

The simulator takes as input the description of a quantum network in terms of pseudocode (a text-formatted file).
Quantum networks are drawn by making use of a graphical user interface.
For this purpose, we developed Quantum Computer Circuit Editor (QCCE), a Microsoft Windows application.
QCCE contains graphical symbols for each of the qubit operations described in Section~\ref{QC}
and for the quantum Fourier transform. Examples of quantum networks drawn by QCCE are shown in Fig.~\ref{adder}
and Fig.~\ref{QFT}. QCCE also contains an interpreter that takes the quantum network as input and
generates a text-formatted file that contains pseudocode describing all operations to be performed on the qubits, including
the swap operations described in Section~\ref{PARA1}. An example of the pseudocode is given in appendix~\ref{pseudo}.
Since drawing a quantum network for a large quantum computer can be quite cumbersome,
it is often more efficient to write a dedicated program that generates the pseudocode language directly.
Since the quantum computer simulator software is a standalone application that takes ASCII files as input,
any software package (such as \url{http://www.phys.cs.is.nagoya-u.ac.jp/~watanabe/qcad/index.html})
that can draw quantum networks may be used, as long as it generates
the input files (including the swap operations) in the format that the simulator can understand.

\begin{table}[t]
\caption{Overview of the computer systems used for testing the parallel
quantum computer software.
Simulations have been performed on machines located at
SARA Computing and Networking Services, Amsterdam, The Netherlands (SARA);
Forschungszentrum J\"ulich, J\"ulich, Germany (FZ-J);
Computer Center, University of Groningen, Groningen, The Netherlands (RuG);
The Institute for Solid State Physics (ISSP) of the University of Tokyo, Tokyo, Japan (ISSP);
Cray Inc., Seattle, USA (Cray Inc.).
}
\begin{center}
\begin{ruledtabular}
\begin{tabular}{cccccc}
\noalign{\vskip 2pt}
& SGI Altix 3700 & IBM Regatta p690+ & IBM Blue Gene/L & Hitachi SR11000/J1  & Cray X1E\\
\hline
\noalign{\vskip 4pt}
 Location  &  SARA  &  FZ-J  &  RuG  &  ISSP    &  CRAY Inc.\\
 Memory  &  832 GB &  5.2 TB  &  3.1 TB  & 2.8 TB    &  512 GB\\
 \# CPUs  &  416  &  1312   &  12288   & 2048     &  128 \\
 CPU type & Intel Itanium 2 & Power 4+ & Power PC 970 & POWER5 & 64-bit Cray X1E MSP\\
 CPU clock& 1.3 GHz & 1.7 GHz & 0.7 GHz & 1.9 GHz & 1.13 GHz\\
\end{tabular}
\end{ruledtabular}
\label{SDMPC}
\end{center}
\end{table}

\section{Simulation results}\label{RESULTS}

In this section we present and discuss the results of running the
massive parallel quantum computer simulator on various parallel computers.

The computers on which we perform the simulations and their characteristics are
listed in Table~\ref{SDMPC}.
On all computers we have tested pure OpenMP, pure MPI and combined MPI/OpenMP code.
In practice it turns out that the OpenMP code is up to a factor of two slower
than the pure MPI code.
As a consequence, also the MPI/OpenMP code is slower than the pure MPI code.
A detailed study of the (dis)advantages of this hybrid approach in the case of optimized parallel code
for specific gates will be presented in a future publication~\cite{TRIE06}.
In what follows we only present the results of the pure MPI code.
Because of the special architecture of the Cray X1E processors (Cray Multistreaming Processors (MSPs)
with various vector pipes per MSP), some CRAY specific directives have been added to the code.

A rather simple algorithm to test the scaling properties of the simulation software with the number of qubits is
to perform a Hadamard operation on each qubit of a $L$-qubit quantum computer.
However, simulating gates for single-, two- and three-qubit operations only is not sufficient to test the correctness
of quantum computer simulation software and to compare the performance of various massive parallel computers.
Hence, for this purpose, we consider some more sophisticated quantum algorithms.
A first nontrivial example is the qubit adder that adds the content of several qubit registers.
This quantum algorithm is built up from several quantum operations, including a quantum Fourier transform,
and has the advantage that it is very easy to check the correctness of the simulation result.


\begin{figure*}[t]
\begin{center}
\includegraphics[width=18cm]{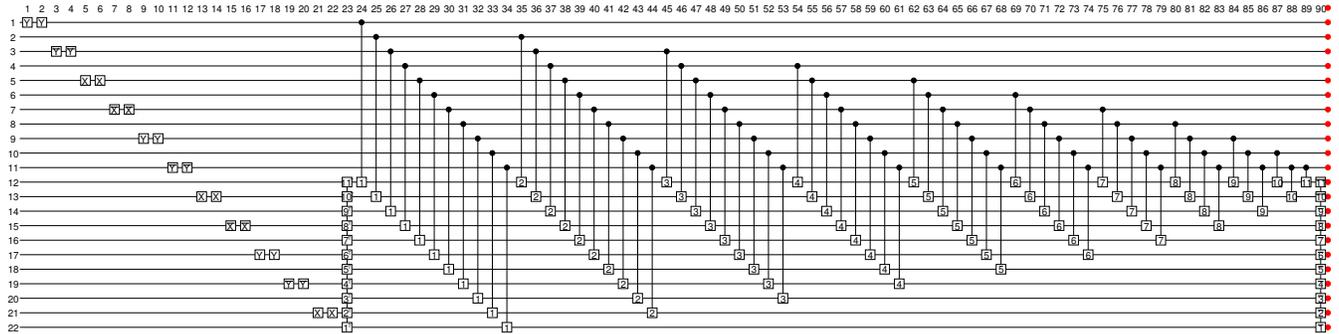}
\end{center}
\caption{Quantum network to add the content of two 11-qubit registers modulo $2^{11}$.
The horizontal lines, numbered from 1 to 22, denote the qubits involved in the quantum operations.
The horizontal row of numbers labels the quantum operations.
The first 22 quantum operations are the single-qubit operations $XX$, $YY$, ${\BAR X}{\BAR X}$
and ${\BAR Y}{\BAR Y}$ that are used to put numbers (in binary notation) in the registers.
The first register (qubits 1 to 11) contains the number 1365 and the second one (qubits 12 to 22) contains the
number 682. Quantum operation 23 is a quantum Fourier transform on the second register.
The full quantum network for this quantum Fourier transform is depicted in Fig.~\ref{QFT}.
Quantum operations 24-89 are controlled phase shift operations that add the information
contained in the first register to the content of the second one.
Quantum operation 90 is a inverse quantum Fourier transform.
The (red) dots at the end of each horizontal line denote the measurement process at each qubit.
}
\label{adder}
\end{figure*}

\begin{figure*}[t]
\begin{center}
\includegraphics[width=18cm]{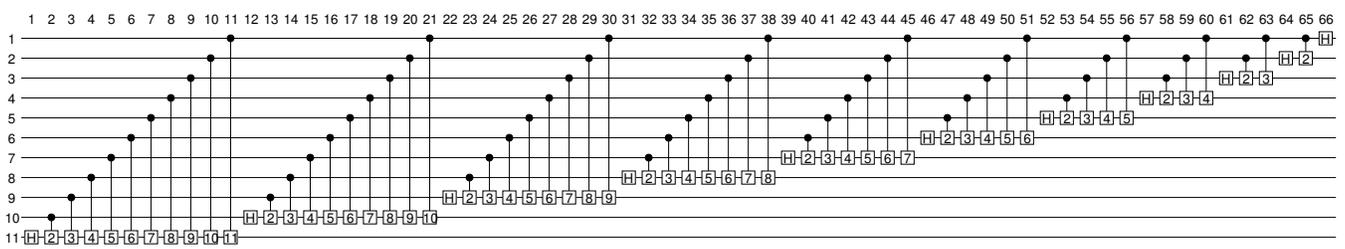}
\end{center}
\caption{Quantum network, built from Hadamard gates and single-qubit phase shifts,
to perform a quantum Fourier transform on eleven qubits~\cite{NIEL00}.
}
\label{QFT}
\end{figure*}

Figure \ref{adder} shows the quantum network to add the content of two
eleven-qubit registers. Note that here the qubits are numbered starting from one.
We use a similar modular structure as was used for adding the content of three
four-qubit registers~\cite{BETT02}.
The basic idea of the algorithm is to use the Quantum Fourier Transform (QFT) to first transfer
the information in one register to the phase factors of the amplitudes, then use controlled phase shifts
to add the information from the other registers, and finally QFT back to the original representation.
These quantum networks perform addition modulo two to the power of the
number of qubits in the register on which the QFT is being applied.

Originally all qubits are in the state $\KET{0}$. The single-qubit operations $XX$, $YY$, ${\BAR X}{\BAR X}$
and ${\BAR Y}{\BAR Y}$ are used to put numbers (in binary notation) in the registers.
Note that all these operations bring a qubit in the state $\KET{1}$. Hence, in principle we could use only
one of these operations to put numbers in the registers. However, in order to test the various single-qubit
operations we used the four different operations.
In the example shown in Fig.~\ref{adder} the first register (qubits 1 to 11) contains the number 1365
and the second register (qubits 12 to 22) the number 682. We perform a QFT on the second register
to transfer the information in the second
register to the phase factors of the amplitudes. The full quantum network for a QFT on eleven qubits is shown in
Fig.~\ref{QFT}. After applying the QFT to the second register, we use controlled phase shifts to add the information from
the first register to the content of the second one. Finally we QFT back to the original
representation.

\begin{figure*}[t]
\begin{center}
\includegraphics[width=12cm]{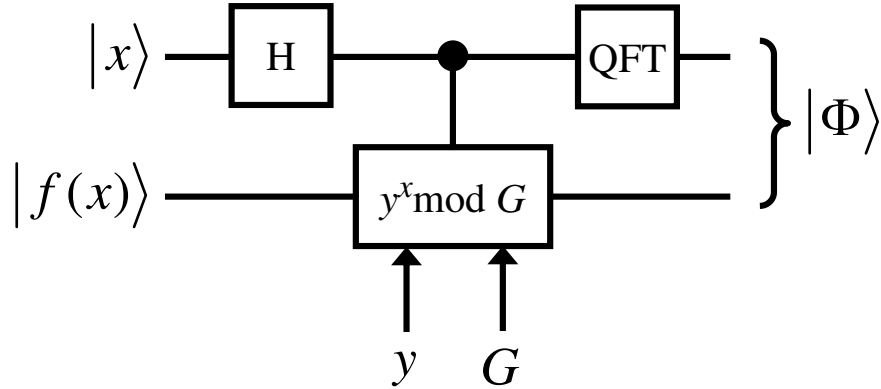}
\end{center}
\caption{Schematic diagram of Shor's algorithm.
}
\label{Shor}
\end{figure*}

\begin{table}[t]
\caption{Representative results of running Shor's algorithm on the massive parallel quantum computer simulator.
The total number of qubits of the simulator is given by $L$ and $X$ is the number of qubits in the $x$-register.
The remaining $L-X$ qubits are used to hold the results of $y^x\mod G$, where $G$ is the composite
integer to be factorized. The values of $1<y<G$ are chosen such that the period $r$ is even and
$y^{r/2}\neq\pm 1\mod G$.
The rows labeled by $\EXPECT{Q_i}$ for $i=0,...,X-1$ give the simulation results for the expectation values of
the $X$ qubits.
These results are the same as those obtained from the analytical expression Eq.~\Eq{Shor7}.
$L=24$: Thinkpad T43P (number of MPI processes $N=1$, elapsed time $t_E=41$ s);
$L=33$: SGI Altix 3700 ($N=64$, $t_E=630$ s), IBM Regatta p690+ ($N=64$, $t_E=810$ s);
$L=36$: IBM Regatta p690+ ($N=512$, $t_E=970$ s) and IBM Blue Gene/L (N=4096, $t_E=170$ s).
}
\begin{center}
\begin{ruledtabular}
\begin{tabular}{cccccc}
$L$       &$24$ &   $33$              & $33$              &  $36$             &  $36$                 \\
$X$       &$16$ &   $22$              & $22$              &  $24$             &  $24$                 \\
$G$       &$247=13\times19$&   $1961=37\times53$ & $2047=23\times89$ &  $4087=61\times67$&  $4033=37\times109$   \\
$y$       &$194$           &   $1698$            & $617$             &  $1392$           &  $1693$               \\
\hline
\noalign{\vskip 2pt}
$r$       &   $18$            &   $468$             & $88$              &  $44$             &  $108$\\
\hline
\noalign{\vskip 2pt}
$\EXPECT{Q_0}$     & $0.500$ &   $ 0.500 $         & $ 0.500$          &  $ 0.500$         &  $ 0.500$             \\
$\EXPECT{Q_1}$     & $0.500$ &   $ 0.500 $         & $ 0.500$          &  $ 0.500$         &  $ 0.500$             \\
$\EXPECT{Q_2}$     & $0.500$ &   $ 0.500 $         & $ 0.500$          &  $ 0.500$         &  $ 0.500$             \\
$\EXPECT{Q_3}$     & $0.445$ &   $ 0.500 $         & $ 0.461$          &  $ 0.461$         &  $ 0.500$             \\
$\EXPECT{Q_4}$     & $0.445$ &   $ 0.500 $         & $ 0.459$          &  $ 0.459$         &  $ 0.490$             \\
$\EXPECT{Q_5}$     & $0.445$ &   $ 0.500 $         & $ 0.455$          &  $ 0.455$         &  $ 0.482$             \\
$\EXPECT{Q_6}$     & $0.444$ &   $ 0.499 $         & $ 0.455$          &  $ 0.455$         &  $ 0.482$             \\
$\EXPECT{Q_7}$     & $0.444$ &   $ 0.496 $         & $ 0.455$          &  $ 0.455$         &  $ 0.482$             \\
$\EXPECT{Q_8}$     & $0.444$ &   $ 0.496 $         & $ 0.455$          &  $ 0.455$         &  $ 0.482$             \\
$\EXPECT{Q_9}$     & $0.444$ &   $ 0.496 $         & $ 0.455$          &  $ 0.455$         &  $ 0.481$             \\
$\EXPECT{Q_{10}}$  & $0.444$ &   $ 0.496 $         & $ 0.455$          &  $ 0.455$         &  $ 0.481$             \\
$\EXPECT{Q_{11}}$  & $0.444$ &   $ 0.496 $         & $ 0.455$          &  $ 0.455$         &  $ 0.481$             \\
$\EXPECT{Q_{12}}$  & $0.444$ &   $ 0.496 $         & $ 0.455$          &  $ 0.455$         &  $ 0.481$             \\
$\EXPECT{Q_{13}}$  & $0.444$ &   $ 0.496 $         & $ 0.455$          &  $ 0.455$         &  $ 0.481$             \\
$\EXPECT{Q_{14}}$  & $0.444$ &   $ 0.496 $         & $ 0.455$          &  $ 0.455$         &  $ 0.481$             \\
$\EXPECT{Q_{15}}$  & $0.500$ &   $ 0.496 $         & $ 0.455$          &  $ 0.455$         &  $ 0.481$             \\
$\EXPECT{Q_{16}}$  &         &   $ 0.496 $         & $ 0.455$          &  $ 0.455$         &  $ 0.481$             \\
$\EXPECT{Q_{17}}$  &         &   $ 0.496 $         & $ 0.455$          &  $ 0.455$         &  $ 0.481$             \\
$\EXPECT{Q_{18}}$  &         &   $ 0.496 $         & $ 0.455$          &  $ 0.455$         &  $ 0.481$             \\
$\EXPECT{Q_{19}}$  &         &   $ 0.496 $         & $ 0.500$          &  $ 0.455$         &  $ 0.481$             \\
$\EXPECT{Q_{20}}$  &         &   $ 0.500 $         & $ 0.500$          &  $ 0.455$         &  $ 0.481$             \\
$\EXPECT{Q_{21}}$  &         &   $ 0.500 $         & $ 0.500$          &  $ 0.455$         &  $ 0.481$             \\
$\EXPECT{Q_{22}}$  &         &                     &                   &  $ 0.500$         &  $ 0.500$             \\
$\EXPECT{Q_{23}}$  &         &                     &                   &  $ 0.500$         &  $ 0.500$             \\
\noalign{\vskip 2pt}
\end{tabular}
\end{ruledtabular}
\label{Shortab}
\end{center}
\end{table}

As another nontrivial test we implement Shor's algorithm on a quantum
computer containing up to 36 qubits. Briefly, Shor's algorithm finds the prime factors $p$ and $q$ of a composite
integer $G=p\times q$ by determining the period of the function $f(x)=y^x\mod G$ for $x=0,1,\ldots $ Here, $1< y< G$
should be coprime (greatest common divisor of $y$ and $G$ is 1) to $G$ (otherwise, since $G=p\times q$, $y=p$ or $y=q$, so we
already guessed the solution). Let $r$ denote the period of $f(x)$, that is $f(x)=f(x+r)$.
If the chosen value of $y$ yields an odd period $r$, we repeat the algorithm with another choice for $y$,
until we find an $r$ that is even. Once we have found an even period $r$, we compute $y^{r/2}\mod G$.
If $y^{r/2}\neq \pm 1\mod G$, then we find the factors of $G$ by calculating the greatest common divisors of
$y^{r/2}\pm 1$ and $G$.

On an ideal quantum computer, this algorithm can be carried out with a number of unitary operations that increases as a
polynomial, not as the exponential, of the number of qubits required to store the number $G$.
The schematic diagram of Shor's algorithm is shown in Fig.~\ref{Shor}. The quantum computer has $L$ qubits.
There are two qubit registers: A $x$-register with $X$ qubits
to hold the values of $x$ and a $f$-register with $F=L-X$ qubits to hold the values of $f(x)=y^x\mod G$.
What is the largest number we can factorize on a quantum computer with $L$ qubits?
If we write the number $G$ in its binary representation $G=\sum_{i=0}^{g_1-1} 2^in_i$, where $n_i=0,1$, then
it is clearly seen that $F=g_1$.
For Shor's algorithm to work properly, that is to find the correct period $r$ of $f(x)$,
the number of qubits $X$ in the $x$-register should satisfy~\cite{SHOR99}
\begin{equation}
G^2\leq 2^{X}\leq 2G^2.
\label{Shor0}
\end{equation}
Writing $G^2=\sum_{i=0}^{g_2-1} 2^im_i$, where $m_i=0,1$, it follows from Eq.~\Eq{Shor0} that $g_2-1\leq X\leq g_2+1$.
Omitting numbers $G$ that can be written as a power of two (which are trivial to factorize), the smallest value
for $X$ (and hence the largest value for $F$) is given by $g_2$. Hence, $L=g_1+g_2$. Since either $g_2=2g_1-1$
or $g_2=2g_1$, it follows that the maximum number of qubits that can be reserved for the $f$-register is give by
$F=g_1=\lfloor (L+1)/3\rfloor$.
For example, on a 36-qubit quantum computer $G=4087=61\times 67\leq 2^{12}$ is the largest integer that
can be factorized by Shor's algorithm.

The initial state of the machine is
$\KET{\Phi_0}=\KET{0}$. After applying Hadamard operations to all qubits of the $x$-register we have
\begin{equation}
\KET{\Phi_1}=2^{-X/2}\sum_{x=0}^{2^X -1}\KET{x}_x\KET{0}_f,
\label{Shor1}
\end{equation}
where $\KET{.}_x$ and $\KET{.}_f$ denote the state of the $x$ and $f$-register, respectively.
Then, we compute $f(x)$ for each value $x=0,\ldots ,X-1$. This is implemented as a conditional operation, as
indicated by the middle box in Fig.~\ref{Shor}.
After the second step, the quantum computer is in the state
\begin{equation}
\KET{\Phi_2}=2^{-X/2}\sum_{x=0}^{2^X-1}\KET{x}_x\KET{f(x)}_f.
\label{Shor2}
\end{equation}
The period $r$ of $f(x)$ can now be determined by applying a Fourier transformation, not to $\KET{f(x)}_f$ but to
$\KET{x}_x$, as indicated in Fig.~\ref{Shor}. To see how this step works, we use the periodicity of
$f(x)$ to rewrite Eq.~\Eq{Shor2} as
\begin{equation}
2^{-X/2}\sum_{x=0}^{2^X-1}\KET{x}_x\KET{f(x)}_f=2^{-X/2}\sum_{x=0}^{r-1}(\KET{x}_x+\KET{x+r}_x+\ldots )\KET{f(x)}_f.
\label{Shor3}
\end{equation}
Using the Fourier representation of $\KET{x}_x$ we obtain
\begin{eqnarray}
\KET{\Phi_3}&=&2^{-X/2}\sum_{x=0}^{2^X-1}\KET{x}_x\KET{f(x)}_f\nonumber\\
&=&2^{-X}\sum_{k=0}^{2^X-1}\sum_{x=0}^{r-1}
e^{2\pi ikx/2^X}\left ( 1+e^{2\pi ikr/2^X}+e^{4\pi ikr/2^X}+\ldots +e^{2\pi ikr(s-1)/2^X}\right )
\KET{k}_x\KET{f(x)}_f\nonumber \\
&+&2^{-X}\sum_{k=0}^{2^X-1}\sum_{x=0}^{s-1}e^{2\pi ikx/2^X}e^{2\pi ikrs/2^X}\KET{k}_x\KET{f(x)}_f,
\end{eqnarray}
where $s=\lfloor 2^X/r \rfloor$ denotes the largest integer $s$ such that $rs\leq 2^X$.
The probability to observe the quantum computer in the state $\KET{k}$ reads
\begin{equation}
p_k(r)=\frac{r}{2^{2X}}\left(\frac{\sin (\pi krs/2^X)}{\sin (\pi kr/2^X)}\right)^2-\frac{2^X-rs}{2^{2X}}
\frac{\sin (\pi kr(2s+1)/2^X)}{\sin (\pi kr/2^X)}.
\label{Shor5}
\end{equation}
The function $p_k(r)$ is strongly peaked for all $kr\approx 2^X$. The probability to observe the machine
in the state $\KET{k}$ for which $kr\approx 2^X$ is $p_k(r)\approx r^{-1}$~\cite{EKER96}.
Given the observed state $\KET{k}$, we can find $r$
because the condition Eq.~\Eq{Shor0} guarantees that there exists exactly one function $k^{\prime}/r$ that
satisfies~\cite{SHOR99}
\begin{equation}
\left| \frac{k}{2^X}-\frac{k^{\prime}}{r}\right|\leq \frac{1}{2^{X+1}}.
\label{Shor6}
\end{equation}
The fraction $k^{\prime}/r$ (with $r<G$) can be found effectively by computing the convergents of the continued fraction
representation of $k/2^{X}$~\cite{HARD00,SHOR99,EKER96,NIEL00}.
From Eq.~\Eq{Shor5} we can easily compute the expectation values of the qubits of the $x$-register.
Let $Q_i\KET{x_i}=x_i\KET{x_i}$ for $x_i=0,1$, then
\begin{eqnarray}
\EXPECT{Q_i}&=&\BRA{\Phi_3}Q_i\KET{\Phi_3}=\sum_{k,k^{\prime}}\BRACKET{\Phi_3}{k^{\prime}}\BRA{k^{\prime}}Q_i\KET{k}
\BRACKET{k}{\Phi_3}\nonumber\\
&=&\sum_{k,k^{\prime}}\BRACKET{\Phi_3}{k^{\prime}}\delta_{k,k^{\prime}}\BRACKET{k}{\Phi_3}\BRA{k}Q_i\KET{k}\nonumber\\
&=&\sum_k p_k(r)k_i,
\label{Shor7}
\end{eqnarray}
where $k_i$ denotes the $i$th bit of $k$. The quantum computer simulator should reproduce the values of all
$\EXPECT{Q_i}$, for $i=0,\ldots , X-1$, otherwise there is definitely something wrong in the simulation.
On the other hand, agreement is not a guarantee that the simulator is free of errors. Obviously, for a $L=36$ qubit
simulation, storing the final state for further analysis requires a lot ($>$ 1TB) of disk space.
To alleviate this problem, we have added to the simulator a procedure that takes the final state as input and generates
a user-specified number of basis states $\KET{x}$ with probability $|\BRACKET{x}{\Phi_3 (x)}|^2$.
This procedure is also fully parallel and uses all available MPI processes.

In Table~\ref{Shortab} we show the results of running Shor's algorithm on the massive parallel computer simulator.
The example for $L=24$ qubits was run on an IBM Thinkpad T43P.
The two examples for $L=33$ qubits were both run on the SGI Altix 3700 and on the IBM Regatta p690+.
The two examples for $L=36$ are both run on the IBM Regatta p690+ and on the
IBM Blue Gene/L.
All the tests desribed in this section confirm that the simulator is producing correct results.

\section{Benchmark results}\label{BENCHMARKS}

\begin{figure*}[t]
\begin{center}
\includegraphics[width=12cm]{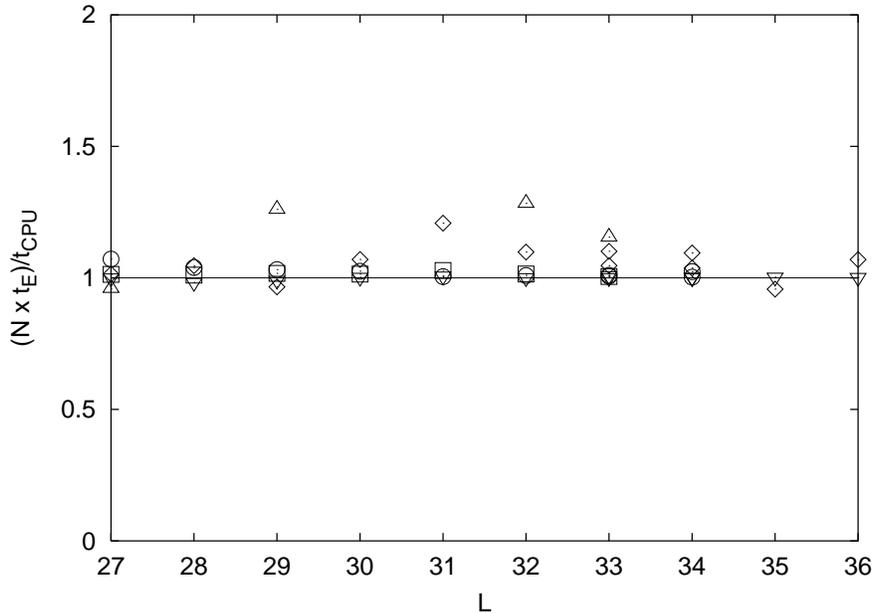}
\end{center}
\caption{Elapsed time $t_E$ divided by the CPU time $t_{CPU}$ and multiplied by the number of MPI processes $N$
as a function of the number of qubits $L$.
Diamonds: IBM Regatta p690+; squares: SGI Altix 3700; flipped triangles: IBM Blue Gene/L;
circles: Hitachi SR11000/J1; triangles: Cray X1E.
The horizontal line at $N\times t_E/t_{CPU}=1$ shows the ideal case
in which the exponential increase of the problem size is exactly compensated by the exponential increase
in the number of processors and memory size. Deviations from the ideal case are due to interprocessor communication.
}
\label{elapstime}
\end{figure*}

\begin{figure*}[t]
\begin{center}
\includegraphics[width=12cm]{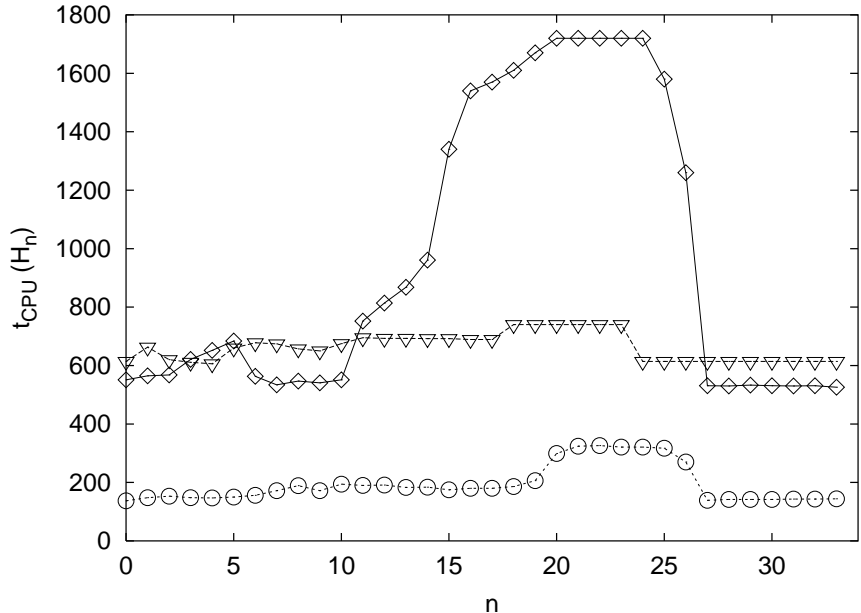}
\end{center}
\caption{CPU time $t_{CPU}$ for a Hadamard operation $H_n$ carried out on qubit $n$ of a quantum computer with
$L=34$ qubits ($0<n\leq 33$).
Diamonds: IBM Regatta p690+; flipped triangles: IBM Blue Gene/L; circles: Hitachi SR11000/J1.
The lines are guides to the eye.
For $15\leq n\leq 26$, the increase of CPU time on the IBM Regatta p690+ is due to less appropriate
memory access and can be reduced by appropriate optimization techniques~\cite{TRIE06}.
}
\label{CPUtimeHadamard}
\end{figure*}

\begin{figure*}[t]
\begin{center}
\includegraphics[width=12cm]{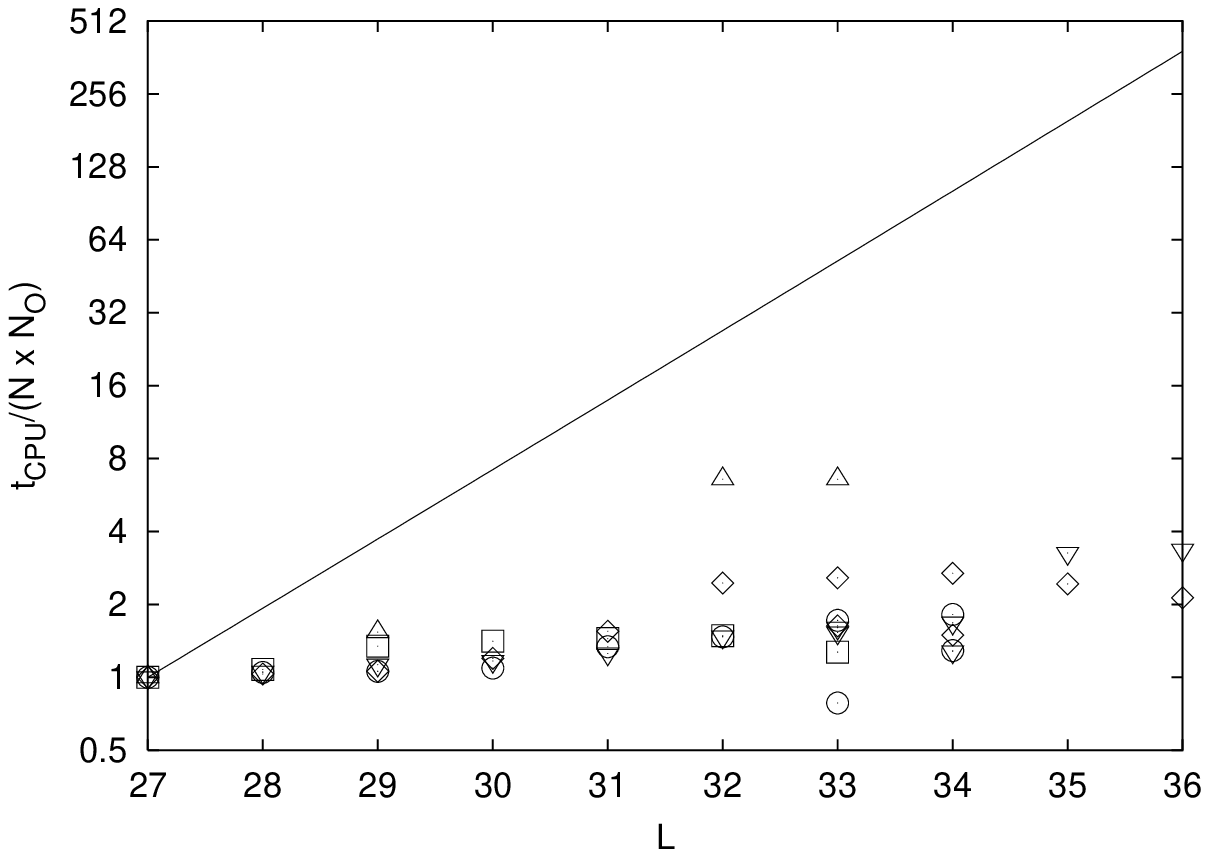}
\end{center}
\caption{CPU time $t_{CPU}$ divided by the number of quantum operations $N_O$
multiplied by the number of MPI processes $N$
as a function of the number of qubits $L$.
Diamonds: IBM Regatta p690+; squares: SGI Altix 3700; flipped triangles: IBM Blue Gene/L;
circles: Hitachi SR11000/J1; triangles: Cray X1E.
The line is given by $27\times 2^{L-27}/N_O$.
}
\label{CPUtime}
\end{figure*}

The quantum algorithms that we use for benchmarking the simulator software
are the Hadamard operation performed on each qubit,
qubit adders to add the content of three 11-qubit registers, two 17-qubit registers, five 7-qubit registers
and three 12-qubit registers.
Tables~\ref{Regatta}-\ref{Cray} of appendix~\ref{Simres} summarize the results on the various machines.
In order to compare the performance of the different machines we use the data
from Tables~\ref{Regatta}-\ref{Cray} and plot them in Figs.~\ref{elapstime}-\ref{CPUtime}.
Figure \ref{elapstime} depicts the elapsed time $t_E$ divided by the CPU time $t_{CPU}$ and multiplied by the
number of MPI processes $N$ as a function of the number of qubits $L$.
All data listed in Tables~\ref{Regatta}-\ref{Cray} is processed.
In the ideal case, we expect that $N\times t_E/t_{CPU}=1$, independent of $L$.
As can be seen from Fig.~\ref{elapstime}, on most machines we observe scaling properties that are very close to ideal.
Deviations from the ideal behavior, if there are any, are relatively small.
The IBM Regatta p690+ and the Cray X1E show the largest deviations.

In Fig.~\ref{CPUtimeHadamard} we show the CPU time for a Hadamard operation $H_n$ carried out on qubit $n$
of a quantum computer with $L=34$ qubits.
In this case the CPU time is measured per gate and does not include the time for measuring the qubit.
Recall that we number the qubits starting from zero, so that $0<n\leq 33$.
We only show results for the IBM Regatta p690+ (diamonds), IBM Blue Gene/L (flipped triangles) and the Hitachi SR11000/J1 (circles).
The number of MPI processes on the IBM Regatta p690+, the IBM Blue Gene/L and the Hitachi SR11000/J1 are $N=128$, $N=1024$
and $N=128$, respectively.
Hence, on the IBM Regatta p690+ and the Hitachi SR11000/J1 qubits $0\leq n\leq 26$ are local and
on the IBM Blue Gene/L qubits $0\leq n\leq 23$ are local.
In all cases, the CPU time for $H_n$ carried out on a nonlocal qubit is equal to the CPU time for
the Hadamard gate performed on qubit zero, because according to the algorithm described in Section~\ref{PARA1},
we interchange the nonlocal qubit with the local qubit zero.
On the IBM Regatta p690+, the CPU time to carry out $H_n$ on a local qubit $n$ can differ by a factor
of three depending on the qubit number $n$.
This is a result of how the local memory addresses are accessed
and is thus due to the architecture of the machine.
Techniques to speed up the memory access will be discussed in a future application~\cite{TRIE06}.
On the IBM Blue Gene/L and on the Hitachi SR11000/J1 we only see a relatively
small increase in CPU time if we operate $H_n$ on qubits with inreasing number $n$.
The difference in behavior between the IBM Regatta p690+ and the two other machines is also
reflected in Fig.~\ref{elapstime}, although there it is much less clear.
Figure \ref{CPUtimeHadamard} clearly shows that the Hitachi SR11000/J1 is significantly faster than the IBM Regatta p690+
and the IBM Blue Gene/L.

Figure \ref{CPUtime} shows $t_{CPU}/(N\times N_O)$ as a function of $L$,
$N_O$ being the number of quantum operations (not including swap commands).
All data listed in Tables~\ref{Regatta}-\ref{Cray} is processed.
On all machines, the number of MPI processes $N=2^{L-27}$, except on the IBM Blue Gene/L, $N=8\times 2^{L-27}$.
Therefore, for a program that parallelizes 100\% we expect that $t_{CPU}=\alpha N_O2^{L-27}$, where $\alpha$ is a constant.
Hence, in the ideal case $t_{CPU}/(N\times N_O)$ should be equal to $\alpha$.
From Fig.~\ref{CPUtime} it can be seen that for all machines, $t_{CPU}/(N\times N_O)$ is slightly increasing
as a function of $L$. The deviation from the ideal performance is due to the communication
between various MPI processes. The corresponding increase in CPU time
seems to be the largest for the IBM Regatta p690+ and the Cray X1E.
However, the overall scaling properties of the simulation software are extremely good. For comparison we depict the
line, given by $27\times 2^{L-27}/N_O$, in Fig.~\ref{CPUtime}. This line expresses
the expected behavior of $t_{CPU}/(N\times N_O)$ as a function of $L$ on a non-parallel computer,
that is a computer for which $N=1$. Clearly, with the available hardware and the present simulation software,
it is possible to beat the exponential growth of the simulation problem by simply increasing the memory and
the number of CPUs by a factor of two for each qubit added, at least up to 36 qubits.

From Tables~\ref{Regatta}-\ref{Cray} it is clear that from the point of view of the user,
that is in terms of elapsed time, the IBM Blue Gene/L is the fastest computer.
If we make a comparison based on the CPU time then the Cray X1E is the fastest machine,
followed by the Hitachi SR11000/J1.
We emphasize that, except for the Cray X1E, we have used the same source code
on all machines, that is we did not make an effort to adapt the code
for a particular machine.

Out of curiosity and to demonstrate the portability of the parallel simulation code,
we also test the software on a cluster of three IBM Thinkpad T43P notebooks and one IBM Think Centre A50P PC,
using MPICH2 under Windows XP.
Communication between the machines is handled by a 100 Mbit Ethernet US Robotics router.
In Table~\ref{PC} we show the performance results for carrying out a Hadamard operation on each qubit.
Explicit measurement of the
time spent for communication between the machines shows that for the example with $L=27$ qubits, about
half of $t_E$ and $t_{CPU}$ is spent on network communication.

\section{Discussion}\label{SUMM}

The parallel quantum computer simulator presented in this paper
is an application that shows nearly perfect scaling with the number
of CPUs and problem size.
It is a demanding application in that it can use a significant part of the available memory
and CPUs and can put a heavy burden on the communication network.
Therefore, this simulator may find application as a new type of benchmark to assess
the computational power of the new generation of high-end computer systems.

\begin{acknowledgments}
Support from the ``Nederlandse Stichting Nationale Computer Faciliteiten (NCF)'' is gratefully acknowledged.
We thank Gyan Bhanot (IBM Yorktown) for helping us with porting the code to the IBM BlueGene/L,
ASTRON (NL) for providing access to the IBM BlueGene/L and Arnold Meijsters (RuG) for
helping us with running the simulations on the IBM BlueGene/L.
We are grateful to J. Thorbecke (Cray) for performing the benchmarks on the Cray X1E.
We thank W. Frings at FZJ for fruitful discussions on hardware specifics of the IBM p690+.
This work is partially supported by the Japanese Ministry of Internal Affairs and Communications (Soumu-sho).
Finally, we thank the Supercomputer Center, Institute for Solid State Physics,
University of Tokyo for the facilities and the use of Hitachi SR11000/J1.
\end{acknowledgments}

\appendix
\section{Quantum Computer Circuit Editor (QCCE)}\label{pseudo}
In this appendix we discuss the text-formatted file, generated by QCCE, that is used as input for the simulator
if it has to simulate a Hadamard operation performed on each qubit of a 32-qubit quantum computer.
Lines starting with `` ! '' or `` \# '' are considered as comments. We use these symbols to add comments
to the following text generated by QCCE:

\begin{verbatim}
QUBITS 32                         ! The command QUBITS sets the size of the quantum computer
INITIAL STATE 0                   ! The initial state of the quantum computer is set to |0...0>
MPIPROCESSES 32                   ! Number of MPI processes is set to 32
                                  ! (not used by the OpenMP code)
H 0                               ! Hadamard operation on qubit 0
.                                 ! These lines are omitted here but contain commands for
.                                 ! Hadamard operations carried out on qubits 1-25
.
H 26                              ! Hadamard operation on qubit 26
SWAP 1 0 27                       ! Command to swap 1 pair of qubits: qubits 0 and 27
H 27                              ! Hadamard operation on qubit 27
SWAP 1 27 28                      ! Command to swap qubits 27 and 28
H 28                              ! Hadamard operation on qubit 28
SWAP 1 28 29                      ! Command to swap qubits 28 and 29
H 29                              ! Hadamard operation on qubit 29
SWAP 1 29 30                      ! Command to swap qubits 29 and 30
H 30                              ! Hadamard operation on qubit 30
SWAP 1 30 31                      ! Command to swap qubits 30 and 31
H 31                              ! Hadamard operation on qubit 31
BEGIN MEASUREMENT
DO MEASUREMENT 1 2 3 4 5 6 7 8 9 10 11 12 13 14 15 16 17 18 19 20 21 22 23 24 25 26 31
                                  ! Measurement operation on the listed qubits
SWAP 5 31 1 2 3 4 0 27 28 29 30   ! Command to swap 5 pairs of qubits: (31,0), (1,27), (2,28),
                                  ! (3,29), and (4,30)
DO MEASUREMENT 0 27 28 29 30      ! Measurement operation on the listed qubits
END MEASUREMENT
\end{verbatim}

As can be seen from the example, the general format of a command consists of a keyword followed by some values.
Note that the swap command is not the same as the quantum swap operation. The swap command carries out the
qubit interchange described in Section~\ref{PARA1}.

\section{Simulation results}\label{Simres}

\begin{table}[ht]
\caption{Performance results for the IBM Regatta p690+ for the simulation of various quantum algorithms on
a $L$-qubit ideal quantum computer. $N$: Number of MPI processes; Memory: Memory required to store $\KET{\Phi}$;
$N_O$: Number of quantum operations (not including swap commands);
$t_E$: Elapsed time; $t_{CPU}$: CPU time; H: Hadamard operation; $n$-$m$QBA:$\sum_{i=1}^nx_i$:
Qubit adder to add the content of $n$ $m$-qubit registers. The $n$ registers contain each one number $x$ represented in
binary notation by $m$ bits.
The CPU time and the elapsed time include the time for measuring each qubit.
}
\begin{center}
\begin{ruledtabular}
\begin{tabular}{ccccccc}
\noalign{\vskip 2pt}
 $L$ &   $N$  &      Memory [GB] & $N_O$ & $t_E$ [s] & $t_{CPU}$ [s]&  Quantum algorithm\\
\hline
 27  &     1  &    2     &   27  &  144  &      142  &  H\\
 28  &     2  &    4     &   28  &  159  &      304  &  H\\
 29  &     4  &    8     &   29  &  156  &      646  &  H\\
 30  &     8  &   16     &   30  &  202  &     1510  &  H\\
 31  &    16  &   32     &   31  &  305  &     4040  &  H\\
 32  &    32  &   64     &   32  &  453  &    13200  &  H\\
 33  &    64  &  128     &   33  &  492  &    28600  &  H\\
 34  &   128  &  256     &   34  &  526  &    61500  &  H\\
 33  &    64  &  128     &  286  & 2568  &   157000  &  3-11QBA: 292+585+1170\\
 34  &   128  &  256     &  493  & 4019  &   496000  &  2-17QBA: 26214+104857\\
 35  &   256  &  512     &  194  & 2374  &   635000  &  5-7QBA: 7+9+19+35+65\\
 36  &   512  & 1024     &  342  & 4095  &  1960000  &  3-12QBA: 781+1054+3296\\
\end{tabular}
\end{ruledtabular}
\label{Regatta}
\end{center}
\end{table}

\begin{table}[ht]
\caption{Same as Table~\ref{Regatta} for the SGI Altix 3700 system.
}
\begin{center}
\begin{ruledtabular}
\begin{tabular}{ccccccc}
\noalign{\vskip 2pt}
 $L$ &   $N$  &      Memory [GB]  & $N_O$ & $t_E$ [s]& $t_{CPU}$ [s]&  Quantum algorithm\\
\hline
 27  &     1  &    2     &   27  &  308  &      304  &  H\\
 28  &     2  &    4     &   28  &  343  &      679  &  H\\
 29  &     4  &    8     &   29  &  446  &     1754  &  H\\
 30  &     8  &   16     &   30  &  483  &     3809  &  H\\
 31  &    16  &   32     &   31  &  518  &     8063  &  H\\
 32  &    32  &   64     &   32  &  543  &    17127  &  H\\
 33  &    64  &  128     &  286  & 4112  &   216772  &  3-11QBA: 292+585+1170\\
\end{tabular}
\end{ruledtabular}
\label{SGI}
\end{center}
\end{table}

\begin{table}[ht]
\caption{Same as Table~\ref{Regatta} for the IBM Blue Gene/L.
}
\begin{center}
\begin{ruledtabular}
\begin{tabular}{ccccccc}
\noalign{\vskip 2pt}
 $L$ &   $N$  &      Memory [GB]  & $N_O$ & $t_E$ [s]& $t_{CPU}$ [s]&  Quantum algorithm\\
\hline
 27  &     8  &    2     &   27  &   40  &      320  &  H\\
 28  &    16  &    4     &   28  &   43  &      700  &  H\\
 29  &    32  &    8     &   29  &   48  &     1550  &  H\\
 30  &    64  &   16     &   30  &   52  &     3320  &  H\\
 31  &   128  &   32     &   31  &   58  &     7370  &  H\\
 32  &   256  &   64     &   32  &   70  &    17900  &  H\\
 33  &   512  &  128     &   33  &   79  &    40100  &  H\\
 34  &  1024  &  256     &   34  &   85  &    86800  &  H\\
 33  &   512  &  128     &  286  &  649  &   332000  &  3-11QBA: 292+585+1170\\
 34  &  1024  &  256     &  493  &  934  &   956000  &  2-17QBA: 26214+104857\\
 35  &  2048  &  512     &  194  &  940  &  1920000  &  5-7QBA: 7+9+19+35+65\\
 36  &  4096  & 1024     &  342  & 1707  &  6980000  &  3-12QBA: 781+1054+3296\\
\end{tabular}
\end{ruledtabular}
\label{BG}
\end{center}
\end{table}

\begin{table}[ht]
\caption{Same as Table~\ref{Regatta} for the Hitachi SR11000/J1.
}
\begin{center}
\begin{ruledtabular}
\begin{tabular}{ccccccc}
\noalign{\vskip 2pt}
 $L$ &   $N$  &      Memory [GB]   & $N_O$ & $t_E$ [s]& $t_{CPU}$ [s]&  Quantum algorithm\\
\hline
 27  &     1  &    2     &   27  &   75  &     70    &  H\\
 28  &     2  &    4     &   28  &   79  &    152    &  H\\
 29  &     4  &    8     &   29  &   82  &    318    &  H\\
 30  &     8  &   16     &   30  &   87  &    679    &  H\\
 31  &    16  &   32     &   31  &  108  &   1720    &  H\\
 32  &    32  &   64     &   32  &  123  &   3900    &  H\\
 33  &    64  &  128     &   33  &  148  &   9410    &  H\\
 34  &   128  &  256     &   34  &  164  &  20500    &  H\\
 33  &   128  &  128     &  286  &  586  &  74300    &  3-11QBA: 292+585+1170\\
 34  &   128  &  256     &  493  & 1653  & 211000    &  2-17QBA: 26214+104857\\
\end{tabular}
\end{ruledtabular}
\label{Hitachi}
\end{center}
\end{table}

\begin{table}[ht]
\caption{Same as Table~\ref{Regatta} for the Cray X1E.
}
\begin{center}
\begin{ruledtabular}
\begin{tabular}{ccccccc}
\noalign{\vskip 2pt}
 $L$ &   $N$  &      Memory [GB]   & $N_O$ & $t_E$ [s]& $t_{CPU}$ [s]&  Quantum algorithm\\
\hline
\noalign{\vskip 4pt}
 27  &     4  &    2     &   27  &   12  &       50  &  H\\
 29  &     8  &    8     &   29  &   26  &      165  &  H\\
 32  &    32  &   64     &   32  &  125  &    3117   &  H\\
 33  &    64  &  128     &  286  & 1006  &   55750   &  3-11QBA: 292+585+1170\\
\end{tabular}
\end{ruledtabular}
\label{Cray}
\end{center}
\end{table}

\begin{table}[ht]
\caption{Same as Table~\ref{Regatta} for a cluster of three IBM Thinkpad T43P notebooks (each having 1 GB of memory
and a 2.13 GHz CPU) and
one IBM Think Centre A50P PC (having 2 GB of memory and a 2.8 GHz CPU), using MPICH2 under
Windows XP.
Communication between the machines is handled by a 100 Mbit Ethernet US Robotics router. Explicit measurement of the
time spent for communication between the machines shows that for the example with $L=27$ qubits
half of $t_E$ and $t_{CPU}$ is spent on the swap operation.
}
\begin{center}
\begin{ruledtabular}
\begin{tabular}{ccccccc}
\noalign{\vskip 2pt}
 $L$ &   $N$  &      Memory [GB]   & $N_O$ & $t_E$ [s]& $t_{CPU}$ [s]&  Quantum algorithm\\
\hline
\noalign{\vskip 4pt}
 26  &     1  &    1     &   26  &   73  &       73  &  H\\
 26  &     4  &    1     &   26  &   81  &      290  &  H\\
 27  &     4  &    2     &   27  &  304  &     1220  &  H\\
\end{tabular}
\end{ruledtabular}
\label{PC}
\end{center}
\end{table}


\clearpage
\pagebreak[4]
\raggedright

\end{document}